\RequirePackage{fix-cm}
\documentclass[smallextended]{svjour3}       
\usepackage{graphicx}
\usepackage{lineno}
\usepackage{natbib}
\usepackage{color}

\journalname{Journal of Artificial Intelligence and Consciousness}

\begin{document}

\newcommand{\sh}[1]{\textcolor{blue}{\textbf{Shimon:} #1}}
\newcommand{\am}[1]{\textcolor{red}{\textbf{Aman: #1}}}

\title{Functionally Effective Conscious AI Without Suffering}

\titlerunning{Ethics of Conscious AI}        

\author{Aman Agarwal         \and
        Shimon Edelman 
}

\authorrunning{Agarwal \& Edelman} 

\institute{A.~Agarwal \at
              Dept.\ of Computer Science \\
              Cornell University \\
              \email{aa2398@cornell.edu}           
           \and
           S.~Edelman \at
                Dept.\ of Psychology \\
                Cornell University \\
                \email{se37@cornell.edu} 
}

\date{}

\maketitle

\begin{abstract}
    Insofar as consciousness has a functional role in facilitating learning and behavioral control, the builders of autonomous AI systems are likely to attempt to incorporate it into their designs. The extensive literature on the ethics of AI is concerned with ensuring that AI systems, and especially autonomous conscious ones, behave ethically. In contrast, our focus here is on the rarely discussed complementary aspect of engineering conscious AI: how to avoid condemning such systems, for whose creation we would be solely responsible, to unavoidable suffering brought about by phenomenal self-consciousness. We outline two complementary approaches to this problem, one motivated by a philosophical analysis of the phenomenal self, and the other by certain computational concepts in reinforcement learning. 
\end{abstract}

\section{The two sides of AI and ethics}
\label{intro}

With the growing presence of autonomous software and devices in daily life, the ethics of Artificial Intelligence (AI) has rightly become an intensely researched and debated topic \citep[e.g.,][]{McCulloch56ethical,Metzinger13,Dignum18,Kuipers19,Floridi19,JobinEtAl19}. Most of that research and debate is focused on ensuring that autonomous AI systems behave ethically --- that is, in accordance to certain human ideals (as opposed to actual human behavior, which we, rather understandably given the history of our species, would not want AI to emulate). To put it concisely and rather bluntly, we do not wish, ever, to suffer at the hands of the machines that we create. 

It is easy to see, however, that this ethical concern is asymmetrical and that its complement --- the possibility of the machines suffering at our hands --- should receive at least as much attention. Indeed, it should probably receive more attention: because any AI would owe its very existence to \emph{us}, our share of ethical responsibility in this entire matter is far larger. Crucially, this concern only applies if the AI that we create or cause to emerge becomes conscious and thereby capable of suffering. In this paper, we examine the nature of the relevant kind of conscious experience, the potential functional reasons for endowing an AI with the capacity for feeling and therefore for suffering, and some of the possible ways of retaining the functional advantages of consciousness, whatever they are, while avoiding the attendant suffering.

What \emph{is} suffering? Thomas \citet[p.244]{Metzinger17} has recently offered the following qualified take:

\begin{quote}
We lack a comprehensive theory of conscious suffering. One of the key desiderata is a conceptually convincing and empirically plausible model of this very specific class of phenomenal states: those that we do \emph{not} want to experience if we have any choice, those states of consciousness which folk-psychology describes as ``suffering.''
\end{quote}

\noindent
On this approach, the central characteristic of suffering is a loss of autonomy and of cognitive control, possibly signifying an impending or ongoing physical damage to the body. This insight serves as a bridge between, on the one hand, the phenomenal nature of suffering, as well as of conscious awareness in general, and, on the other hand, the functional roles of consciousness. One of these roles is plausibly held to be centralized control, such as facilitated by the ``global workspace'' postulated by some theories of consciousness \citep{Baars88,Shanahan10,DehaeneEtAl14}. Another role is facilitating learning \citep{Cleeremans11,CleeremansEtAl20}, especially of the unsupervised and autonomous variety \citep{Metzinger17}. Importantly, for all this to matter to suffering, any such information-processing role must be accompanied by obligatory ``caring'' about learning and behavioral outcomes. Indeed, the inseparability of awareness from feelings and affect has been postulated in \citep[e.g.,][]{Merker07,Metzinger17,MoyalFeketeEdelman20}). The question thus arises whether or not sufficiently effective learning and control, as well as generally good behavioral outcomes, can be achieved by a system that is neither entirely devoid of phenomenality, nor given to unavoidable suffering. 

The remainder of this paper is structured as follows. In section~\ref{sec:theory} we briefly survey theories of consciousness that have a bearing on the nature of suffering, notably the concept of phenomenal self-model (PSM) as developed in the recent work of Metzinger. Section~\ref{sec:function} then takes up the question of the possible functional role(s) of consciousness, which leads naturally to computational considerations of effective learning and behavior regulation. Section~\ref{sec:nosuff} applies lessons from the preceding discussion in an attempt to determine whether or not consciousness without suffering is feasible and if yes, whether such consciousness can still fulfill the relevant functional needs. Section~\ref{sec:comp} examines computational approaches to implementing functionally effective yet suffering-free conscious systems. Finally, section~\ref{sec:discussion} concludes the paper with a brief summary.

\section{The nature of suffering and its relation to conscious experience in general}
\label{sec:theory}

The question of the nature of suffering, as distinguished from its ethical dimensions, is rarely, if ever, raised in theoretical treatments of consciousness --- a peculiar omission, which prompted \citet{Metzinger17} to refer to suffering as ``a cognitive scotoma.'' Insofar as suffering involves negative affect, it should in principle fall within the scope of any theoretical account of conscious phenomenal experience. In other words, a theory of qualia must be at the same time a theory of affect, for the simple reason that qualia, or feelings, do as a rule incorporate affective dimensions \citep[e.g.,][]{Havermans11,Krieglmeyer10,KrieglmeyerEtAl13,BeattyEtAl16,EderEtAl16,TurnerEtAl17}. In practice, however, popular theories of consciousness, such as the Global Workspace Theory \citep[e.g.,][]{DehaeneEtAl14} or the Information Integration Theory \citep{OizumiAlbantakisTononi14}, stop short of addressing the question of the nature of affect, and therefore of suffering. The same goes for higher-order theories (HOT) of consciousness, as reviewed, for instance, in \protect\citep{Rosenthal09}; these offer accounts of pain, but do not seem to mention suffering.

The psychology of affect has been usefully summarized by \citet[p.31]{Panksepp05}: ``Affect is the subjective experiential-feeling component that is very hard to describe verbally, but there are a variety of distinct affects, some linked more critically to bodily events (homeostatic drives like hunger and thirst), others to external stimuli (taste, touch, etc.). Emotional affects are closely linked to internal brain action states, triggered typically by environmental events. All are complex intrinsic functions of the brain, which are triggered by perceptions and become experientially refined. Psychologists have traditionally conceptualized such ``spooky'' mental issues in terms of valence (various feelings of goodness and badness --- positive and negative affects), arousal (how intense are the feelings), and surgency or power (how much does a certain feeling fill one's mental life). There are a large number of such affective states of consciousness, presumably reflecting different types of global neurodynamics within the brain and body.'' 

For our present purposes, the valence dimension of affect is of most interest: without negative affective states there would be no suffering. Suffering is, however, more than just negative affect. As \citet[p.244]{Metzinger17} notes, suffering is a class of phenomenal states that ``we do \emph{not} want to experience if we have any choice.'' In other words, suffering is a state of negative affect from which the sufferer cannot escape by simply wishing it away. As we shall see in section~\ref{sec:function}, the stress on inescapability in this formulation makes explicit the intimate connection between the experiential flavor of suffering and its presumed evolutionary-functional role. It also serves to distinguish between the first-person experience of suffering and the suffering of others, which is not directly felt. Ethical theorists have argued that the latter should be as objectionable to oneself as the former. According to \citet[p.160]{Nagel86}, for instance, ``the pain can be detached in thought from the fact that it is mine without losing any of its dreadfulness\dots\ suffering is a bad thing, period, and not just for the sufferer\dots\ This \emph{experience} ought not to go on, \emph{whoever} is having it.'' \citet[p.135]{Parfit11} quotes from Nagel and concurs with his moral stance. Our concern here is, however, exclusively with suffering as it presents itself to the sufferer, rather than with the ethical problems that it creates for others. Even if pain, as Nagel puts it, ``can be detached \emph{in thought} [our emphasis] from the fact that it is mine'', it is \textit{a priori} unclear whether or not it can be so detached \emph{in lived experience}. To address this crucial question, we turn to Metzinger's analysis of phenomenal experience.

Briefly, \citet{Metzinger00} develops a representationalist account of the first-person perspective, centered on the phenomenal self-model (PSM): a ``multimodal representational structure, the contents of which form the contents of the consciously experienced self.'' Crucially, the PSM is generally phenomenally transparent (the T~condition), i.e., it is normally not recognized as merely representational by the system itself.\footnote{The T~condition can be illustrated by contrasting the normal dream state, during which the the dreamer does not realize he or she is dreaming (transparent PSM), with lucid dreaming, during which the PSM becomes opaque and the dreamer may even be able to exert control over the dreamt universe.} The contents of the PSM include the phenomenal properties of ``mineness,'' selfhood, and perspectivalness. According to \citet{Metzinger17}, the PSM is an ``instrument for global self-control,'' and is therefore fundamental to the phenomenology of suffering, which is characterized by a loss of control in addition to negative valence (the NV condition). This analysis motivates our proposed strategy for avoiding suffering as a matter of direct experience, which we describe in section~\ref{sec:nosuff}. 




\section{The possible functional benefits of endowing AI with consciousness}
\label{sec:function}

Given that phenomenal consciousness as we know it incorporates affective dimensions (see the references in section~\ref{sec:theory}), being conscious sets the agent up for suffering. The simplest way to avoid suffering would then be to give up phenomenal consciousness itself. For an ethically minded engineer, this translates into an imperative to stick to information processing architectures that, to the best of our understanding, cannot result in artificial consciousness. According to the Information Integration Theory, for instance, feedforward network architectures (``zombie networks'') are incapable of supporting consciousness \citep[Fig.20]{OizumiAlbantakisTononi14}. The Geometric Theory \citep{FeketeEdelman11} and its successor, the Dynamical Emergence Theory \citep{MoyalFeketeEdelman20}, hold that systems that are devoid of properly structured intrinsic dynamics are likewise devoid of phenomenality.

Restricting robotics to the building of artificial ``zombies'' is not, however, a viable engineering option if consciousness confers any significant functional advantages for an AI system or robot. In a commercial setting, technologies that promise to be more effective displace less effective ones even if this comes at the price of serious ethical flaws, and AI is not exempt from this tendency. We therefore next turn to the question of the functional benefits of consciousness. This question is seldom addressed in consciousness research, perhaps because it is taken for granted that the benefit is essentially cognitive in the narrow sense, stemming from the ``global'' access to information that consciousness affords (as per the Global Workspace theory, mentioned earlier). This default account may be compared to to the ``radical plasticity'' thesis of \citet{Cleeremans11}, according to which learning to care is the central component as well as the functional benefit of emergent consciousness.

\citet[p.252]{Metzinger17} goes further down this road by assuming that not just consciousness but specifically suffering is a prerequisite for autonomy: ``[\dots] functionally speaking, suffering is necessary for autonomous self-motivation and the emergence of truly intelligent behaviour.'' In an evolutionary setting, this assumption makes intuitive sense insofar as (i) reinforcement learning is universally employed by living systems in honing adaptive behavior, and (ii) an autonomous system by definition must provide its own source of drive, as per the principle of intrinsic motivation \citep{Barto13}. Furthermore, evolutionary simulations suggest that performance-driven positive affect alone is not as effective in motivating an agent as an alternation of positive and negative affective states, brought about, respectively, by successes and failures \citep{GaoEdelman16a}; moreover, such a balance between happiness and unhappiness can serve as an effective intrinsic motivator \citep{GaoEdelman16b}. If it were possible for the agent to \emph{choose} not to experience negative affect, suffering would be avoided, but the question still remains whether or not the price for that would be failing to learn quickly and well from the consequences of behavior. 

Reinforcement learning is not only an evolutionary-biological universal, but also the method of choice in an engineering setting. While RL was shown to be effective in certain types of tasks (notably, games; \citealp{SilverEtAl16,VinyalsEtAl19}), its use across tasks and in unconstrained real-world situations is limited by the extreme difficulty of formulating good universally applicable reward functions. One remedy for this is the inverse RL approach, in which the development goal is not to equip the learning system with a ready-made reward function, but rather to let it try to approximate the developers' preferences, choices, and habits, defined over classes of outcomes \citep[e.g.,][]{christiano2017deep}. A more radical approach is to let the system under development learn the reward functions entirely on its own. This, however, would seem to put us back on square one: if autonomy is indeed essential, Metzinger's view that suffering is needed for effective learning would be supported. We  return to this key question in the following section.

\section{Functionally effective consciousness without suffering: first-principles considerations}
\label{sec:nosuff}

If consciousness indeed brings with it unique functional advantages, is it possible to engineer conscious AI systems that would benefit from these, while ensuring that such systems are not thereby doomed to suffer? Following the account in \citet{Metzinger17}, if consciousness itself is retained, logically there are four ways to mitigate suffering: (a) eliminating the PSM, (b) eliminating the NV-condition, (c) eliminating the T-condition, or (d) maximizing the unit of identification (UI). The first three directly follow from our discussion in Section~\ref{sec:theory}, but we argue that they likely do not satisfy our functional needs. On the other hand, the fourth approach is a promising direction, which we will describe and focus on subsequently. 
 
For functionally beneficial consciousness, the mere possibility of experience is not sufficient. The conscious systems must additionally perceive themselves as entities in relationship with their surrounding world, and have a sense of ownership over the arising conscious experiences. In other words, these systems must be \emph{self}-conscious, not merely conscious, i.e., they must activate a phenomenal self model (PSM). Similarly, they must have preferences regarding their experiences, not least so that they prefer the experience of fulfilling desired goals over frustrating them. Stated differently, these systems must be sensitive to the positive or negative valence of phenomenal experiences. Thus, approaches (a) and (b) to eliminating suffering while retaining the functional advantages of being conscious are not feasible. 
 
Next, approach (c) raises the interesting question of whether phenomenal transparency is also necessary for proper functioning. In principle, it might be possible that an active PSM and sensitivity to NV could endure along with their associated functional benefits, even in the absence of transparency. In this situation, the system would lose the naive realism and immediacy that are normally associated with its experiences, by becoming aware of their representational character, and yet, continue to function according to the dictates of the PSM and NV avoidance. However, awareness of the representational character of the contents of consciousness, which means awareness of the increasingly complex stages of information processing behind them, would likely severely hinder the functional efficiency of the conscious machines without providing any valuable actionable information. So, option (c) is also unlikely to work.  
 
The final approach is similar to~(c) in that it also targets the phenomenology of identification with the PSM as an antidote to suffering, but it does so in a seamless fashion making it much more viable for our purposes. \citet{Metzinger18} describes the unit of identification (UI) as that which the system consciously identifies itself with. Ordinarily, when the PSM is transparent, the system identifies with its PSM, and is thus conscious of itself as a \emph{self}. But it is at least a logical possibility that the UI not be limited to the PSM, but be shifted to the ``most general phenomenal property''  \citep{Metzinger17} of \emph{knowing} common to all phenomenality including the sense of self. In this special condition, the typical subject-object duality of experience would dissolve; negatively valenced experiences could still occur, but they would not amount to suffering because the system would no longer be experientially \emph{subject} to them. It is worth noting that such ``non-dual awareness'' which cuts through the ``illusion of the self'' has been the soteriological focus of various spiritual traditions, most notably Buddhism, as the key to liberation from suffering and to enlightenment. Furthermore, this approach also fits nicely with the reductionist view of personal identity put forth by \citet{Parfit84}, who acknowledged its connection to the Buddha's philosophy.

Two questions remain to be addressed. First, how can such maximization of the UI be achieved in machines? Second, can a PSM that is sufficiently effective for functioning be maintained under the maximized UI condition?

\subsection{Realizing no-suffering}

In \citep{Metzinger18}, the concept of the Minimal Phenomenal Experience (MPE) is developed as the most general phenomenal property that underlies all phenomenal experiences, and thus serves as the natural candidate target for UI maximization.\footnote{The apparent conflict in the nomenclature here is resolved by noting that under UI maximization MPE is minimal in the sense of being the least specific.} The MPE is characterized by wakefulness, contentlessness, self-luminosity and a quality of ``knowingness'' without object, which is normally unnoticed but can become available to introspective attention under the right conditions. Intuitively, MPE likely corresponds to the phenomenal state described in Buddhist and Advaita Vedanta philosophies as ``emptiness'' \citep[e.g.,][]{Siderits03,Priest09} and ``witness-consciousness'' \citep[e.g.,][]{Albahari09} respectively, as attested to by highly advanced meditators. Importantly, Metzinger proposes that in the human brain, the MPE is implemented by the Ascending Reticular Activation System (ARAS), which causes auto-activation by which the brain wakes itself up. As the most general signal which the brain must regulate, the ever-present yet contentless ARAS-signal is, arguably, what corresponds to the MPE. That the MPE might have such a stable neural correlate is not surprising if it is indeed fundamental to phenomenal experience \emph{as such}, distinct from any concepts, thoughts etc., appearing in consciousness. 

A critically important point is that all other phenomenal experiences such as the PSM are superimposed onto the MPE, so it should be possible to attend to regular conscious phenomena while simultaneously being aware of the inherent all-encompassing MPE in the background. This motivates our claim that UI maximization (and thus, suffering avoidance) can be achieved in conscious machines by building in their identification with the MPE via both physical design (analogous to hardware) and conceptual/programmatic training (analogous to software). If the physical design of the machines is such that there is a component which performs the analogous function of auto-activation as the ARAS does in humans, then its signal could be tuned to make the MPE salient in the machines. Since a necessary condition for noticing the MPE is knowledge that there \emph{is} such a thing to be noticed, and then paying attention appropriately \citep{Metzinger18}, the machines would then have to be trained to attend to their accessible-by-design MPE. This could be done via practices common in certain types of meditation that encourage ``turning attention upon itself'' and thus realizing that there is no center (or minimal self) from which consciousness is directed (for a review of the relevant meditation techniques, such as Dzogchen, see e.g.\ \citealp{DahlEtAl15}). In addition to training their attention, the machines could also be provided with the relevant conceptual knowledge about the nature of consciousness (such as the cited works of Metzinger and perhaps the present paper).
 
More generally speaking, there is no fundamental reason why the self-other illusion of duality (such as it is) should persist in conscious machines: it is quite possible that it will be easier for these artificial systems to realize the empty, selfless nature of conscious experience than it is for us. After all, once machines attain certain capabilities, they reliably excel at them. With any luck (and of course, given our proposed measures above), that will also be the case for their meditative capabilities.  




\subsection{Effective functioning without suffering}

If a conscious machine does not suffer because it phenomenologically identifies with the MPE, then will it still be able to function effectively and ethically? We have argued above that the (a) PSM and (b) NV avoidance conditions are conducive to proper functioning, while Metzinger leaves open the question of whether or not the PSM condition can be fulfilled under a maximized UI \citep{Metzinger17}. 

We hypothesize that the functional benefits of consciousness can indeed be maintained when the UI is maximized to the MPE. The key idea is that proper functioning relies on \emph{automatic, subpersonal}, but nonetheless conscious processes, as entailed by the physical design of the system; it should be possible for these processes to continue unhindered while the system identifies with the MPE upon which these conscious experiences are necessarily superimposed. In particular, the functionally requisite PSM and NV avoidance conditions can be maintained as subpersonal processes that do not amount to suffering (which is by nature personal) since the system is not identified with the PSM, but with the MPE, which is completely \emph{impersonal}. This hypothesis is supported by the observation that human beings are already subject to automatic subpersonal conscious processes, including thought (i.e., mind wandering), for roughly two-thirds of their lifetimes, and these processes lay the foundation for functionally beneficial reward prediction, delay discounting etc.\ \citep{Metzinger13myth}. Expanding the UI to the MPE would lead to gaining meta-awareness of these ongoing automatic conscious processes, analogous to gaining meta-awareness of the breath or the heartbeat. This enables an escape from suffering, but not from the relentless progress of the processes themselves, analogous to the inescapable biological imperatives of breathing and heartbeat. 





Furthermore, the UI shift to the MPE may even enhance functioning by making the machines more pro-social and ethical, as well as immune to \emph{self}-induced neuroticism. Since the MPE is completely impersonal, identification with it can directly engender a profound sense of fundamental \emph{sameness} with all other conscious beings, and thus naturally lead to pro-social and moral behaviors. Conceptually, we predict that UI maximization to the MPE will lead to a positive ``top-down'' effect on the ongoing subpersonal processes responsible for various behaviors, similar to the beneficial effects reported in human meditators \citep{DahlEtAl15}.





\section{Functionally effective consciousness without suffering: some further computational ideas}
\label{sec:comp}

Let us return to definition of suffering in \citet{Metzinger17}, which posits that an agent suffers when it identifies with a state of negative affect, from which it cannot to escape. In the previous section, we considered the possibility of shifting the agent's self-identification from the affective states to MPE, the minimal phenomenal experience that underlies all conscious states according to Metzinger's analysis. In some sense, this amounts to \emph{restricting} the self. In contrast, we now focus on \emph{expanding} it, in such a manner that the agent identifies not only with the affective states but also with their causal predecessors. The computational framework of reinforcement learning, which we already invoked in section~\ref{sec:function}, offers just the conceptual tools that can be recruited for this purpose. 

First, we note that reinforcement learning appears to be the most effective when it is intrinsically motivated --- that is, when the rewards originate within the agent, as opposed to being supplied from the outside (see \citep{SinghLewisBarto10} for an evolutionary perspective and \citep{BaldassareMirolli13} for a book-length treatment). Second, if the mechanisms of reward are indeed to be contained within the agent, standard considerations of transparent, robust, and effective design require that these mechanisms be kept separate from those that implement actions. The result is the modular actor-critic scheme for RL, in which action selection and reward appear as distinct modules. Importantly, both these modules are part of the agent \citep[see][fig.2]{Barto13}.

As long as the agent's phenomenal self-model, PSM, holds the actor module alone to constitute the self, negative affect brought about by negative reward is inescapable, resulting in suffering. But what if the PSM is modified --- specifically, extended so as to include the critic module? We hypothesize that such an expansion of the self would mitigate suffering, both by ``diluting'' it (through direct realization of the proximate causes of the negative affect) and by opening up to the possibility of eventual cessation of negative affect as progress towards the performance goals set by the critic is observed.\footnote{This move would not, however, alleviate the ``deserved'' suffering brought about by the pursuit of unattainable goals.}

A more radical option with regard to repurposing the PSM calls for shutting it down and only activating it when needed. Assuming that consciousness, and specifically the PSM, serves to facilitate learning (as per section~\ref{sec:function}), the primary need for it arises during the agent's development or during acquisition of additional skills. During routine operation, consciousness in an artificial agent may only be required when particularly difficult behavioral choices need to be made,\footnote{Cf.\ \citet[p.338]{SmithShieldsWashburn03}: ``If you watch an aging cat consider a doubtful leap onto the dryer, you will suspect that what \citet[p.93]{James90} said is true, `Where indecision is great, as before a dangerous leap, consciousness is agonizingly intense'.''} especially under circumstances that threaten the system's integrity --- what we would call life-threatening situations. 

To understand this mode of operation, it is useful to recall Metzinger's \citeyearpar[p.553]{Metzinger03} idea of the conscious brain as a ``total flight simulator'' --- one that simulates not only the environment that is being navigated, but also the pilot, that is, the virtual entity that serves as the system's self. In dreamless sleep, the pilot is not needed and is temporarily shut down. Arguably, an agent can be engineered so that it can continue to function --- in routine situations --- without a PSM (as a variety of philosophical zombie), with ``sentinel'' programs in place that would reconstitute the PSM as needed. While in a zombie state, such an agent would be incapable of suffering.

\section{Summary}
\label{sec:discussion}

We have outlined two classes of approaches to the problem of the proliferation of suffering arising in connection with engineering conscious AI systems. The first approach calls for ensuring that such systems have both the capability and the propensity to identify with an impersonal Minimal Phenomenal Experience, of the kind that human meditators have been employing for centuries in their attempt to alleviate the suffering associated with the presence of the first-person self and the self-world duality. The second, complementary approach involves an attempt to modify the Phenomenal Self-Model, the computational construct that implements the first-person self, so as to break the default connection between dispreferred outcomes and the inescapable negative affect that amounts to suffering. The question remains open whether or not these two approaches can indeed prevent, or at least alleviate, artificially engineered suffering without detracting from the systems' performance. There can, however, be no doubt that we, as the potential creators of conscious AI, are obligated to do everything in our power not to elevate performance over ethical considerations that cut to the very core of existence and phenomenal experience. 

\begin{acknowledgements}
AA would like to acknowledge the work of Sam Harris, especially the \textit{Waking Up} book and podcasts for an invaluable introduction to the nature of consciousness from a first-person perspective, both as a philosophical exercise and a meditation practice. Harris's work also led AA to discover the books \textit{On Having No Head} by Douglas Harding, \textit{I Am That} by Nisargadatta Maharaj, and  \textit{The Flight of the Garuda} by Keith Dowman, which have highly influenced many of the ideas presented in this paper.
\end{acknowledgements}

\bibliographystyle{spbasic_updated}      
\bibliography{AI-and-suffering.bib}   

\begin{thebibliography}{45}
\providecommand{\natexlab}[1]{#1}
\providecommand{\url}[1]{{#1}}
\providecommand{\urlprefix}{URL }
\expandafter\ifx\csname urlstyle\endcsname\relax
  \providecommand{\doi}[1]{DOI~\discretionary{}{}{}#1}\else
  \providecommand{\doi}{DOI~\discretionary{}{}{}\begingroup
  \urlstyle{rm}\Url}\fi
\providecommand{\eprint}[2][]{\url{#2}}

\bibitem[{Albahari(2009)}]{Albahari09}
Albahari M (2009) Witness-consciousness: Its definition, appearance and
  reality. Journal of Consciousness Studies 16:62--84

\bibitem[{Baars(1988)}]{Baars88}
Baars BJ (1988) A cognitive theory of consciousness. Cambridge University
  Press, New York, NY

\bibitem[{Baldassarre and Mirolli(2013)}]{BaldassareMirolli13}
Baldassarre G, Mirolli M (eds)  (2013) Intrinsically Motivated Learning in
  Natural and Artificial Systems. Springer, Berlin,
  \doi{10.1007/978-3-642-32375-1}

\bibitem[{Barto(2013)}]{Barto13}
Barto AG (2013) Intrinsic motivation and reinforcement learning. In:
  Baldassarre G, Mirolli M (eds) Intrinsically Motivated Learning in Natural
  and Artificial Systems, Springer, Berlin, pp 16--47,
  \doi{10.1007/978-3-642-32375-1}

\bibitem[{Beatty et~al.(2016)Beatty, Cranley, Carnaby, and
  Janelle}]{BeattyEtAl16}
Beatty GF, Cranley NM, Carnaby G, Janelle CM (2016) Emotions predictably modify
  response times in the initiation of human motor actions: a meta-analytic
  review. Emotion 16:237--251

\bibitem[{Christiano et~al.(2017)Christiano, Leike, Brown, Martic, Legg, and
  Amodei}]{christiano2017deep}
Christiano P, Leike J, Brown TB, Martic M, Legg S, Amodei D (2017) Deep
  reinforcement learning from human preferences

\bibitem[{Cleeremans(2011)}]{Cleeremans11}
Cleeremans A (2011) The radical plasticity thesis: how the brain learns to be
  conscious. Frontiers in Psychology 2:86

\bibitem[{Cleeremans et~al.(2020)Cleeremans, Achoui, Beauny, Keuninckx, Martin,
  {Mu{\~n}oz-Moldes}, Vuillaume, and {de Heering}}]{CleeremansEtAl20}
Cleeremans A, Achoui D, Beauny A, Keuninckx L, Martin JR, {Mu{\~n}oz-Moldes} S,
  Vuillaume L, {de Heering} A (2020) Learning to be conscious. Trends in
  Cognitive Sciences

\bibitem[{Dahl et~al.(2015)Dahl, Lutz, and Davidson}]{DahlEtAl15}
Dahl CJ, Lutz A, Davidson RJ (2015) Reconstructing and deconstructing the self:
  cognitive mechanisms in meditation practice. Trends in Cognitive Sciences
  19:515--523

\bibitem[{Dehaene et~al.(2014)Dehaene, King, and Marti}]{DehaeneEtAl14}
Dehaene S, King LCJR, Marti S (2014) Toward a computational theory of conscious
  processing. Current Opinion in Neurobiology 25:76--84

\bibitem[{Dignum(2018)}]{Dignum18}
Dignum V (2018) Ethics in artificial intelligence: introduction to the special
  issue. Ethics and Information Technology 20:1--3,
  \doi{10.1007/s10676-018-9450-z}

\bibitem[{Eder et~al.(2016)Eder, Rothermund, and Hommel}]{EderEtAl16}
Eder AB, Rothermund K, Hommel B (2016) Commentary: Contrasting motivational
  orientation and evaluative coding accounts: on the need to differentiate the
  effectors of approach/avoidance responses. Frontiers in Psychology 7:163

\bibitem[{Fekete and Edelman(2011)}]{FeketeEdelman11}
Fekete T, Edelman S (2011) Towards a computational theory of experience.
  Consciousness and Cognition 20:807--827

\bibitem[{Floridi(2019)}]{Floridi19}
Floridi L (2019) Translating principles into practices of digital ethics: Five
  risks of being unethical. Philosophy \& Technology 32:185--193,
  \doi{10.1007/s13347-019-00354-x}

\bibitem[{Gao and Edelman(2016{\natexlab{a}})}]{GaoEdelman16a}
Gao Y, Edelman S (2016{\natexlab{a}}) Between pleasure and contentment:
  evolutionary dynamics of some possible parameters of happiness. PLoS One
  11(5):e0153,193

\bibitem[{Gao and Edelman(2016{\natexlab{b}})}]{GaoEdelman16b}
Gao Y, Edelman S (2016{\natexlab{b}}) Happiness as an intrinsic motivator in
  reinforcement learning. Adaptive Behavior 24:292--305

\bibitem[{Havermans(2011)}]{Havermans11}
Havermans RC (2011) ``{Y}ou say it's liking, {I} say it's wanting\dots’’.
  {O}n the difficulty of disentangling food reward in man. Appetite 57:286--294

\bibitem[{James(1890)}]{James90}
James W (1890) The Principles of Psychology. Holt, New York, available online
  at {http://psychclassics.yorku.ca/James/Principles/}

\bibitem[{Jobin et~al.(2019)Jobin, Ienca, and Vayena}]{JobinEtAl19}
Jobin A, Ienca M, Vayena E (2019) The global landscape of {AI} ethics
  guidelines. Nature Machine Intelligence 1:389--399

\bibitem[{Krieglmeyer et~al.(2010)Krieglmeyer, Deutsch, {De Houwer}, and {De
  Raedt}}]{Krieglmeyer10}
Krieglmeyer R, Deutsch R, {De Houwer} J, {De Raedt} R (2010) Being moved:
  valence activates approach-avoidance behavior independently of evaluation and
  approach-avoidance intentions. Psychological Science 21:607--613

\bibitem[{Krieglmeyer et~al.(2013)Krieglmeyer, {De Houwer}, and
  Deutsch}]{KrieglmeyerEtAl13}
Krieglmeyer R, {De Houwer} J, Deutsch R (2013) On the nature of automatically
  triggered approach-avoidance behavior. Emotion Review 5:280--284

\bibitem[{Kuipers(2019)}]{Kuipers19}
Kuipers B (2019) Perspectives on ethics of ai: Computer science. In: Dubber M,
  Pasquale F, Das S (eds) Oxford Handbook of Ethics of AI, Oxford University
  Press

\bibitem[{McCulloch(1956)}]{McCulloch56ethical}
McCulloch W (1956) Toward some circuitry of ethical robots or an observational
  science of the genesis of social evolution in the mind-like behavior of
  artifacts. Acta Biotheoretica 11:147--156

\bibitem[{Merker(2007)}]{Merker07}
Merker B (2007) Consciousness without a cerebral cortex: a challenge for
  neuroscience and medicine. Behavioral and Brain Sciences 30:63--81

\bibitem[{Metzinger(2000)}]{Metzinger00}
Metzinger T (2000) The *subjectivity* of subjective experience - a
  representationalist analysis of the first-person perspective

\bibitem[{Metzinger(2003)}]{Metzinger03}
Metzinger T (2003) Being No One: The Self-Model Theory of Subjectivity. MIT
  Press, Cambridge, MA

\bibitem[{Metzinger(2013{\natexlab{a}})}]{Metzinger13myth}
Metzinger T (2013{\natexlab{a}}) The myth of cognitive agency: subpersonal
  thinking as a cyclically recurring loss of mental autonomy. Frontiers in
  Psychology 4:931, \doi{10.3389/fpsyg.2013.00931}

\bibitem[{Metzinger(2013{\natexlab{b}})}]{Metzinger13}
Metzinger T (2013{\natexlab{b}}) Two principles for robot ethics. In:
  Hilgendorf E, G{\"u}nther JP (eds) Robotik und Gesetzgebung, Nomos,
  Baden-Baden, Germany, pp 263--302

\bibitem[{Metzinger(2017)}]{Metzinger17}
Metzinger T (2017) Suffering, the cognitive scotoma. In: Almqvist K, Haag A
  (eds) The Return of Consciousness, Axel and Margaret Ax:son Johnson
  Foundation, Stockholm, pp 237--262

\bibitem[{Metzinger and Millière(2020)}]{Metzinger18}
Metzinger T, Millière R (2020) Minimal phenomenal experience: Meditation,
  tonic alertness, and the phenomenology of “pure” consciousness. Radical
  Disruptions of Self-Consciousness

\bibitem[{Moyal et~al.(2020)Moyal, Fekete, and Edelman}]{MoyalFeketeEdelman20}
Moyal R, Fekete T, Edelman S (2020) Dynamical {E}mergence {T}heory ({DET}): a
  computational account of phenomenal consciousness. Minds and Machines
  \doi{10.1007/s11023-020-09516-9}

\bibitem[{Nagel(1986)}]{Nagel86}
Nagel T (1986) The View From Nowhere. Oxford University Press, New York, NY

\bibitem[{Oizumi et~al.(2014)Oizumi, Albantakis, and
  Tononi}]{OizumiAlbantakisTononi14}
Oizumi M, Albantakis L, Tononi G (2014) From the phenomenology to the
  mechanisms of consciousness: {I}ntegrated {I}nformation {T}heory 3.0. PLoS
  Computational Biology 10(5):e1003,588, \doi{10.1371/journal.pcbi.1003588}

\bibitem[{Panksepp(2005)}]{Panksepp05}
Panksepp J (2005) Affective consciousness: Core emotional feelings in animals
  and humans. Consciousness and Cognition 14:30--80

\bibitem[{Parfit(1984)}]{Parfit84}
Parfit D (1984) Reasons and Persons. Clarendon Press, Oxford

\bibitem[{Parfit(2011)}]{Parfit11}
Parfit D (2011) On What Matters. Oxford University Press, Oxford, UK

\bibitem[{Priest(2009)}]{Priest09}
Priest G (2009) The structure of emptiness. Philosophy East \& West 59:467--480

\bibitem[{Rosenthal(2009)}]{Rosenthal09}
Rosenthal DM (2009) Higher-order theories of consciousness. In: Beckermann A,
  McLaughlin BP, Walter S (eds) The Oxford Handbook of Philosophy of Mind,
  Oxford University Press, New York, NY, chap~13, pp 239--252

\bibitem[{Shanahan(2010)}]{Shanahan10}
Shanahan M (2010) Embodiment and the Inner Life. Oxford University Press, New
  York, NY

\bibitem[{Siderits(2003)}]{Siderits03}
Siderits M (2003) On the soteriological significance of emptiness. Contemporary
  Buddhism 4(1):9--23

\bibitem[{Silver et~al.(2016)Silver, Huang, Maddison, Guez, Sifre, {van den
  Driessche}, Schrittwieser, Antonoglou, Panneershelvam, Lanctot, Dieleman,
  Grewe, Nham, Kalchbrenner, Sutskever, Lillicrap, Leach, Kavukcuoglu, Graepel,
  and Hassabis}]{SilverEtAl16}
Silver D, Huang A, Maddison CJ, Guez A, Sifre L, {van den Driessche} G,
  Schrittwieser J, Antonoglou I, Panneershelvam V, Lanctot M, Dieleman S, Grewe
  D, Nham J, Kalchbrenner N, Sutskever I, Lillicrap T, Leach M, Kavukcuoglu K,
  Graepel T, Hassabis D (2016) Mastering the game of {Go} with deep neural
  networks and tree search. Nature 529:484--503

\bibitem[{Singh et~al.(2010)Singh, Lewis, Barto, and Sorg}]{SinghLewisBarto10}
Singh S, Lewis RL, Barto AG, Sorg J (2010) Intrinsically motivated
  reinforcement learning: an evolutionary perspective. IEEE Trans Auton Ment
  Dev 2:70--82

\bibitem[{Smith et~al.(2003)Smith, Shields, and
  Washburn}]{SmithShieldsWashburn03}
Smith JD, Shields WE, Washburn DA (2003) The comparative psychology of
  uncertainty monitoring and metacognition. Behavioral and Brain Sciences
  26:317--373

\bibitem[{Turner et~al.(2017)Turner, Johnston, {de Boer}, Morawetz, and
  Bode}]{TurnerEtAl17}
Turner WF, Johnston P, {de Boer} K, Morawetz C, Bode S (2017) Multivariate
  pattern analysis of event-related potentials predicts the subjective
  relevance of everyday objects. Consciousness and Cognition 55:46--58

\bibitem[{Vinyals et~al.(2019)Vinyals, Babuschkin, Czarnecki, Mathieu, Dudzik,
  Chung, Choi, Powell, Ewalds, Georgiev, Oh, Horgan, Kroiss, Danihelka, Huang,
  Sifre, Cai, Agapiou, Jaderberg, Vezhnevets, Leblond, Pohlen, Dalibard,
  Budden, Sulsky, Molloy, Paine, Gulcehre, Wang, Pfaff, Wu, Ring, Yogatama,
  W{\"u}̈nsch, McKinney, Smith, Schaul, Lillicrap, Kavukcuoglu, Hassabis,
  Apps, and Silver}]{VinyalsEtAl19}
Vinyals O, Babuschkin I, Czarnecki WM, Mathieu M, Dudzik A, Chung J, Choi DH,
  Powell R, Ewalds T, Georgiev P, Oh J, Horgan D, Kroiss M, Danihelka I, Huang
  A, Sifre L, Cai T, Agapiou JP, Jaderberg M, Vezhnevets AS, Leblond R, Pohlen
  T, Dalibard V, Budden D, Sulsky Y, Molloy J, Paine TL, Gulcehre C, Wang Z,
  Pfaff T, Wu Y, Ring R, Yogatama D, W{\"u}̈nsch D, McKinney K, Smith K,
  Schaul T, Lillicrap T, Kavukcuoglu K, Hassabis D, Apps C, Silver D (2019)
  Grandmaster level in {StarCraft} {II} using multi-agent reinforcement
  learning. Nature 575:350--354

\end{thebibliography}

\end{document}